\documentclass[10pt]{article}
\usepackage{graphicx}

\title{Flexible lensless endoscope with a conformationally invariant multi-core fiber}

\author{Viktor Tsvirkun$^{1}$, Siddharth Sivankutty$^{1}$, Karen Baudelle$^{2}$, \\ %
R\'{e}mi Habert$^{2}$, G\'{e}raud Bouwmans$^{2}$, Olivier Vanvincq$^{2}$, \\%
Esben Ravn Andresen$^{1,2,*}$, Herv\'{e} Rigneault$^{1}$}

\date{$^{1}$ Aix Marseille Univ, CNRS, Centrale Marseille, Institut Fresnel, F-13013 Marseille, France\\%
$^{2}$Universit\'{e} de Lille, CNRS UMR 8523 - PhLAM - Laboratoire de Physique des Lasers, Atomes et Mol\'{e}cules, F-59000 Lille, France \\%
$^{*}$ Corresponding author: esben.andresen@univ-lille.fr}

\begin{document}

\maketitle

\begin{abstract}
The lensless endoscope represents the ultimate limit in miniaturization of imaging tools: an image can be transmitted through a (multi-mode or multi-core) fiber by numerical or physical inversion of the fiber's pre-measured transmission matrix. However, the transmission matrix changes completely with only minute conformational changes of the fiber, which has so far limited lensless endoscopes to fibers that must be kept static.
In this letter we report for the first time a lensless endoscope which is exempt from the requirement of static fiber by designing and employing a custom-designed conformationally invariant fiber.
We give experimental and theoretical validations and determine the parameter space over which the invariance is maintained.
\end{abstract}

\section{Introduction} \label{sec:Introduction}

Cellular-level microscopic imaging has long been a vital tool in biomedical research. Recent years have seen numerous efforts to miniaturize imaging instruments to enable cellular-level imaging in behaving animals. 
A recent example is the miniaturized head mounted microscope for fluorescence imaging \cite{ZivCurrOpinNeurobiol2015} (2~g, 9$\times$15$\times$22~mm). This approach has light source, filters, imaging optics, and CMOS camera integrated into a head mounted device. 
A different approach bases the light delivery and collection on optical fiber which brings the advantage that light source and detectors can be remote rather than integrated in the head mounted device. As an added benefit, optical fiber-based approaches can use pulsed laser sources and perform non-linear imaging---to date demonstrations have been using piezo-electric actuators \cite{LombardiniLSA2018} (few grammes, \O{}3~mm $\times$ 40~mm) or micro-electromechanical systems (MEMS) mirrors \cite{ZongNatMeth2017} (2.15~g, \O{}10 $\times$ 40~mm) to perform point-scanning imaging. For an overview of the most "conventional" optical fiber-based approaches, see Ref.~\cite{FlusbergNatMeth2005}.

A new approach to fiber-based endoscopes came about in 2011 \cite{ThompsonOL2011, CizmarOE2011, PapadopoulosOE2012, ChoiPRL2012, AndresenOL2013, AndresenOE2013} which does away with imaging optics between fiber and sample and consequently is often termed "lensless endoscopes". This represents the ultimate limit in miniaturization, since the head-mounted device can be as small as an optical fiber itself. 
Cellular-level imaging in live mice by lensless endoscopes has recently been demonstrated \cite{OhayonBOE2018, Vasquez-LopezLSA2018}, but imaging modality was restricted to one-photon fluorescence imaging and fiber length to few~cm, and the animal subject was fixed. 

The lensless endoscope is most commonly implemented with image acquisition by point-scanning \cite{ThompsonOL2011, OhayonBOE2018, Vasquez-LopezLSA2018, PapadopoulosOE2012, AndresenOL2013}, although other strategies exist \cite{ChoiPRL2012, LoterieOE2015, TsvirkunOL2016}. 
The generic concept of the lensless endoscope is:
(i) The transmission matrix (TM) in the basis of localized modes \cite{PopoffPRL2010} of the fiber is measured in a preliminary step; 
(ii) The columns of the TM are identified as the input fields that give rise to focused output fields;
(iii) The phase masks that convert the laser beam into said input fields are calculated; 
(iv) Said phase masks are displayed sequentially on a spatial light modulator (SLM), resulting in a two-dimensional scan of the output focus, all the while back-scattered fluorescence signal is recorded, as in a point-scanning microscope. 
The fiber can be either a multi-mode fiber (MMF) \cite{OhayonBOE2018, Vasquez-LopezLSA2018, ChoiPRL2012, LoterieOE2015, PloschnerNatPhoton2015} or a multi-core fiber (MCF) \cite{AndresenOL2013, AndresenOE2013, TsvirkunOL2016}. In either case, its transmission matrix $\mathbf{H}$ can be thought of as
\begin{equation}
  \mathbf{H} = \hat{X} \mathbf{H}_{0} 
\end{equation}
where $\mathbf{H}_{0}$ is the pre-measured TM, measured in a reference conformation, and $\hat{X}$ is the "extrinsic contribution" to the TM when the fiber conformation departs from the reference conformation, in general it is represented as an operator \cite{PloschnerNatPhoton2015}. 
In both the MMF and MCF case, the great challenge remains that following each conformational change of the fiber either the additional extrinsic contribution ($\hat X$) or the new TM ($\mathbf{H}$) must be experimentally quantified whether directly \cite{FarahiOE2013, Caravaca-AguirreOE2013} or indirectly \cite{PloschnerNatPhoton2015} in order for aberation-free imaging to continue. This is the main obstacle standing between us and a flexible lensless endoscope which would open the possibility for minimally-invasive imaging in behaving animals.

\section{Results}\label{sec:Results}

\subsection{Twisted multi-core fiber (MCF)} \label{sec:TwistedMCF}
%\subsection{Parameters}
In our earlier work we have shown how lensless endoscopes based on MCF (as opposed to MMF) simplify many of the considerations pertaining to the TM \cite{AndresenJBO2016}. In particular, in Ref.~\cite{TsvirkunOE2017}, we showed that the extrinsic contribution $\hat X$ to the TM of a MCF is simply a diagonal matrix with complex elements of unit norm and argument which is linear in the transverse coordinate, i.e. a matrix with only two free parameters.
In the present article, we take an entirely new approach to overcoming this challenge: We seek to design a MCF for which the extrinsic contribution is identity ($\hat X = \mathbf{I}$) which would render real-time tracking of the TM unnecessary and finally allow lensless endoscopes to reach their full potential as the endoscope fiber could be allowed to flex freely. As we show below a MCF with a twisted geometry can fulfill this requirement.
\begin{figure}[htbp]
	\centering
	\includegraphics[width=\textwidth]{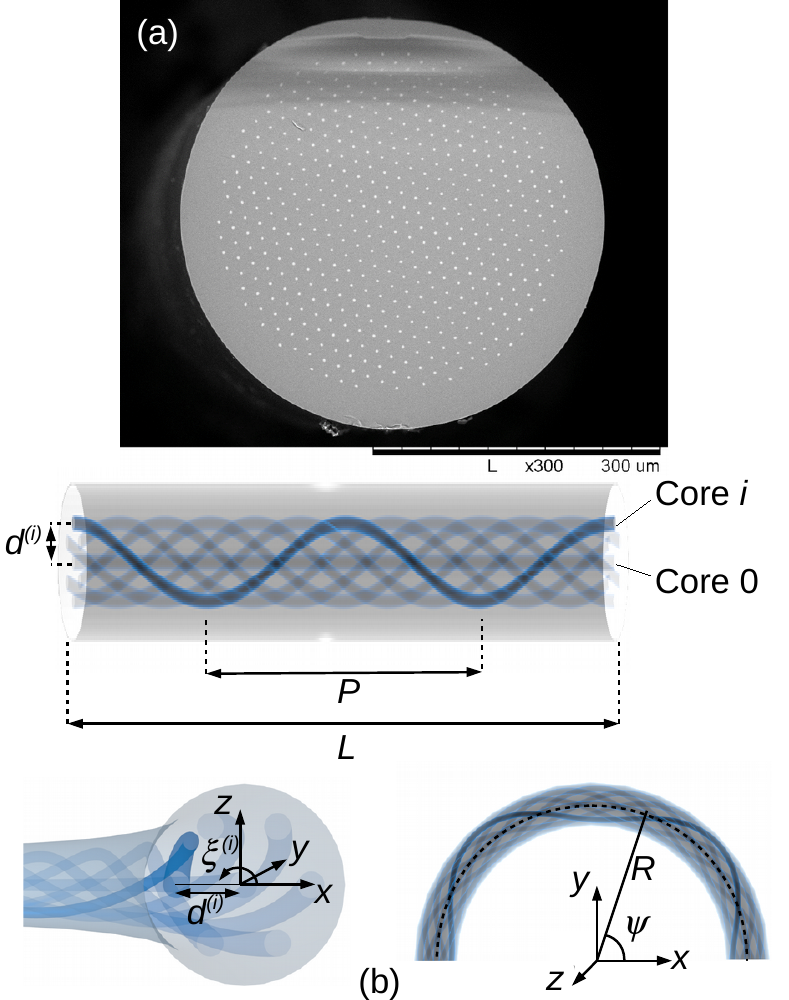}
	\caption{(a) Scanning electron micrograph of the fabricated MCF. Twisted and non-twisted MCFs have the same appearance. (b) Coordinate system and parameter set for the bent, twisted MCF.}
	\label{fig:Fig1}
\end{figure}

\begin{figure}[htbp]
	\centering
	\includegraphics[width = 0.8\textwidth]{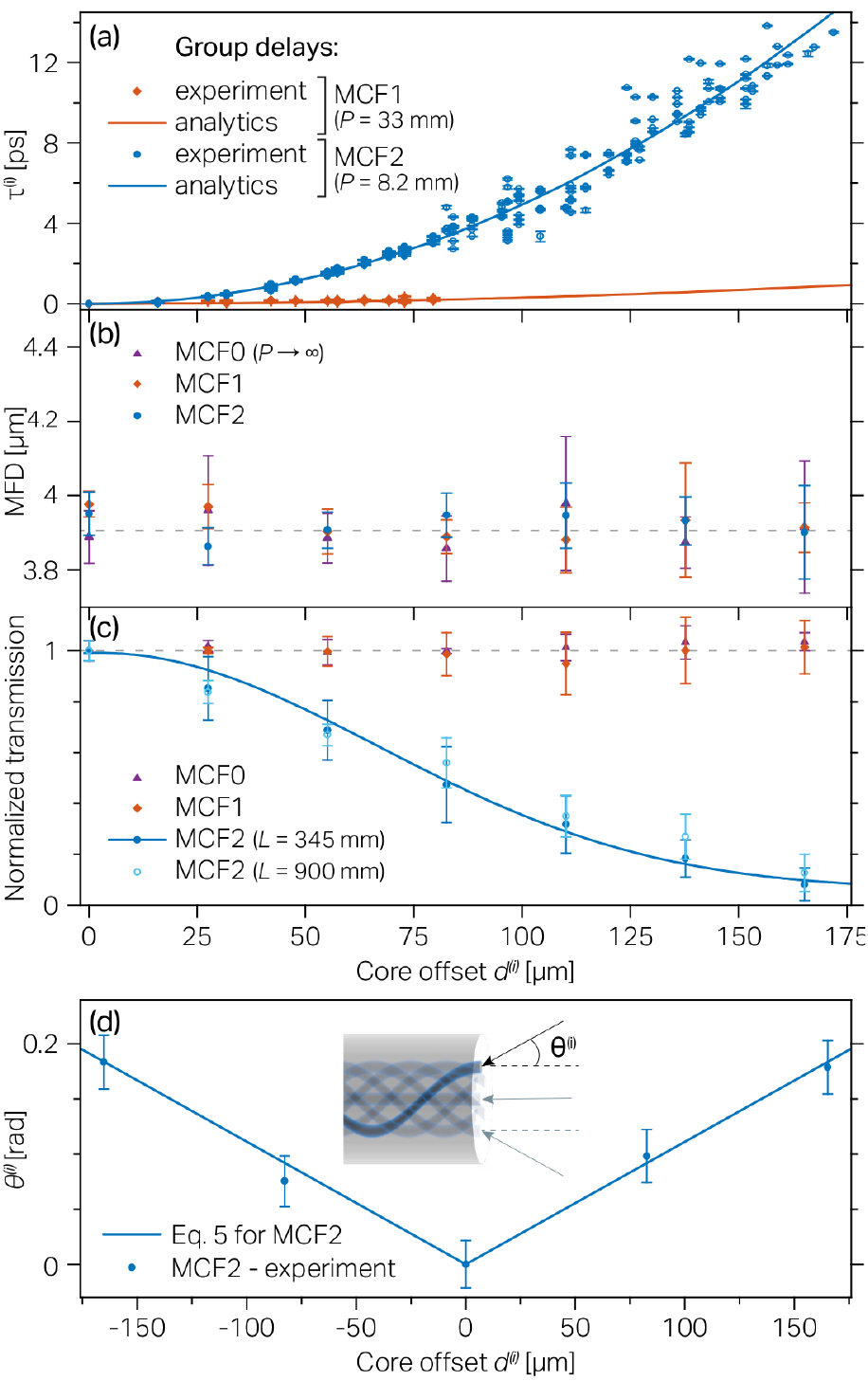}
	\caption{(a) Measured group delays in MCF1 ($P_{1}$~=~33~mm) and MCF2 ($P_{1}$~=~8.2~mm) compared with results of the analytical model.
			(b) Measured mode field diameters in MCF1, MCF2 and non-twisted MCF compared with numerical results.
			(c) Measured attenuation in MCF1, MCF2 and non-twisted MCF0. The solid line is a guide to the eye.
			(d) Measured $\theta^{(i)}$ in MCF2 as a function of $d^{(i)}$ the radial offset of core number $i$. See the inset for the definition of $\theta^{(i)}$.}
	\label{fig:Fig2}
\end{figure}

We designed and fabricated [see the Supplemental Information and Fig.~\ref{fig:Fig1}(a)] two MCFs, "MCF1" and "MCF2" which have the following parameters in common: Number of cores $N$~=~489; core-to-core distance (pitch) $\Lambda$~=~16$\mu$m; parabolic core index profile with index difference relative to the cladding $\Delta n$~=~30$\cdot$10$^{-3}$; core diameter $D$ = 3.5~$\mu$m (single mode).
MCF1 is twisted with a helical period of $P_{1} = 33$~mm, MCF2 with $P_{2} = 8.2$~mm.
See Fig.~\ref{fig:Fig1}(b) for a schematic of the twisted MCFs as well as the coordinates that we will employ.
A non-twisted MCF, "MCF0" with the same parameters but $P_{0} = \infty$ was drawn as well to serve as a reference.

\subsection{Twisted MCF: Intrinsic properties} \label{sec:TwistedMCFInstrinsic}
The theoretical properties of the twisted MCF are derived in the Supplemental Information, a summary is given in the following.
Considering first a straight section of twisted MCF of length $L$.
The center core is unaltered by the twist, the effective index of the fundamental mode in the center core at wavelength $\lambda$~=~1~$\mu$m is $n_{\mathrm{eff}}^{(0)}$~=~1.4626.
Generally, the core $i$, radially offset from the center by $d^{(i)}$, has its physical length modified by the twist as
\begin{equation}
	\label{eq:Li_straight}
	L^{(i)} = \frac{L}{P} \, \sqrt{(2 \pi d^{(i)})^{2} + P^2}.
\end{equation}
As a consequence, the twisted MCF exhibits a radially-dependent native phase delay $\Delta \phi^{(i)}$ as an intrinsic property: 
\begin{equation}
\label{eq:Deltaphii}
\Delta \phi^{(i)} = \phi^{(i)} - \phi^{(0)} = \frac{2 \pi}{\lambda} n_{\mathrm{eff}}^{(0)}  (L^{(i)} - L).
\end{equation}
By the same token, the twisted MCF also exhibits a radially-dependent native group delay $\Delta \tau^{(i)}$ as an intrinsic property
\begin{equation}
	\label{eq:Deltataui}
	\Delta \tau^{(i)} \approx \tau^{(i)} - \tau^{(0)} = n_{\mathrm{eff}}^{(0)} \frac{L^{(i)} - L}{c}. 
\end{equation}
In the strict sense, Eq.~\ref{eq:Deltataui} is approximate, but in the Supplemental Information we demonstrate that group index and refractive index can be used interchangeably without invalidating our conclusions.
Figure~\ref{fig:Fig2}(a) shows the comparison between the native group delays measured in MCF1 and MCF2 by spectral interference \cite{AndresenJOSAB2015} as a function of radial core offset $d^{(i)}$ and those calculated by Eq.~\ref{eq:Deltataui}. We observe excellent agreement. 
Figure~\ref{fig:Fig2}(b) shows experimentally measured mode field diameters compared to those found by numerical simulation. Both show that the twist induces no appreciable variation in the mode field diameter. 
The phase profile of the off-center modes, however, is altered as a consequence of the cores of the twisted MCF no longer being parallel to the MCF axis. This geometrical consideration predicts that the beam exiting core $i$ has an azimuthal component, the angle of the free space beam with the MCF axis being [See Inset in Fig.~\ref{fig:Fig2}(d)]
\begin{equation}
\label{eq:thetai}
\mathrm{sin} [\theta^{(i)}] = n_{\mathrm{eff}}^{(0)} \mathrm{sin} \{ \mathrm{atan} [ \frac{2 \pi d^{(i)}}{P}] \}
\end{equation}
In Fig.~\ref{fig:Fig2}(d) we compare the experimentally measured $\theta^{(i)}$, measured as the optimal coupling angle of the free space beam entering core number $i$, to those predicted by Eq.~\ref{eq:thetai}, and we observe excellent agreement.
Since the off-center cores follow a curved path in space, they are expected to experience loss when $P$ is small. Our numerical simulations showed that no curved path loss ($<$~1~dB/m) due to this effect is expected for neither MCF1 nor MCF2. Curved path loss is expected only for twist period $P~<~2$~mm as detailed in the Supplemental Information. Nevertheless, as seen in Fig.~\ref{fig:Fig2}(c) MCF2 does present a radially-dependent loss, but a cut-back measurement from 900~mm to 345~mm revealed that their origin was not the fiber itself, so we are able to attribute them to coupling loss due to mode mismatch between the input gaussian beam profiles and the MCF modes with non-flat transverse phase cf Eq.~\ref{eq:thetai}. 

\subsection{Twisted MCF: Extrinsic contribution} \label{sec:TwistedMCFExtrinsic}
We consider now a bent section of the twisted MCF with length $L$ and bend radius of curvature $R$ cf Fig.~\ref{fig:Fig1}. Now, in addition to the native phase and group delays, extrinsic contributions to the phase delay $\Delta[\Delta \phi^{(i)}]$ and group delay $\Delta[\Delta \tau^{(i)}]$ appear. 
As we demonstrate in the Supplemental Information, when $L$ = $k \cdot P$, $k \in N$, $L^{(i)}$ is independent of $R$, and as a consequence $\Delta[\Delta \phi^{(i)}] = 0$ and $\Delta[\Delta \tau^{(i)}] = 0$. That is, when the fiber length equals an integer number of twist periods the extrinsic contribution is the identity matrix, and so the MCF is conformationally invariant for all fiber shapes with constant radius of curvature. An intuitive explanation for this is to realize that when the condition is satisfied a given core spends as much time on the exterior of the bent MCF as it does on the interior, so the longitudinal compression of the core on the interior compensated the longitudinal dilation on the exterior.

To demonstrate the conformational invariance, we cut MCF2 to length $L_{2} = 345~\textrm{mm} \approx~42 P_{2}$ and MCF0 to $L_{0}$~=~344~mm and performed a series of imaging experiments.

\subsection{Conformationally invariant endoscope}  \label{sec:BendresilientFocusing}
%\subsection{Point-spread function}

\begin{figure}[htbp]
	\centering
	\includegraphics[width = \textwidth]{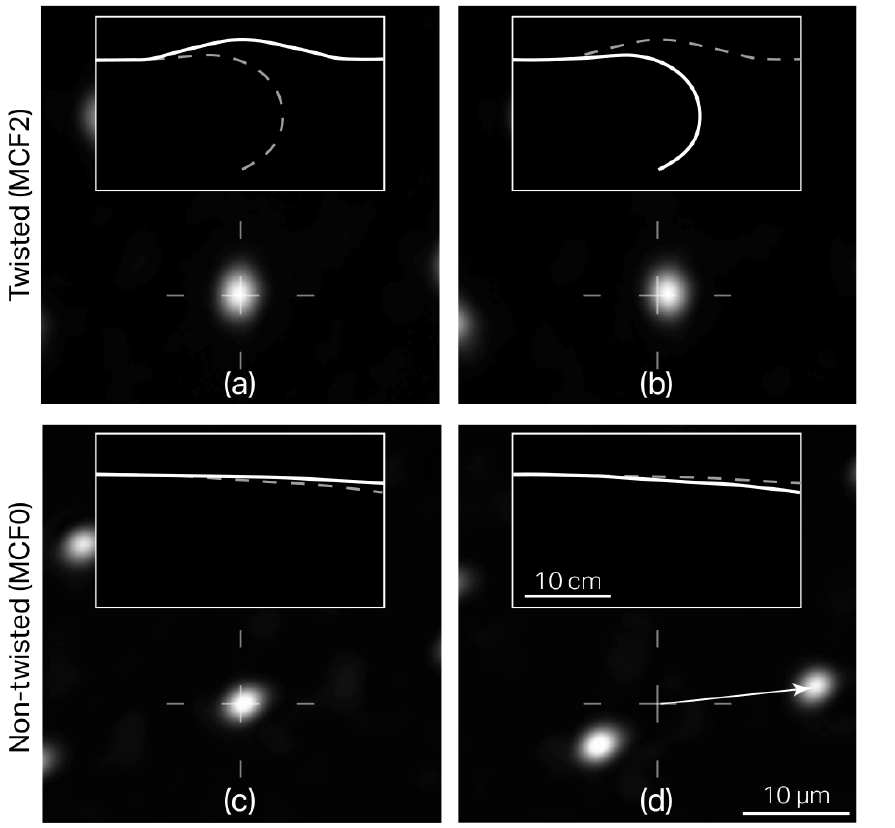}
	\caption{(a) Image of distal focus for the twisted MCF2 ($P$~=~8.2~mm) with $L_{2}=0.345$~m and $R=\infty$; (b) with $R \geq 0.06$~m; see Visualization 1 for the corresponding video.
		(c) Same for the non-twisted MCF0 with $L=0.34$~m and $R=\infty$; (d) with minor bending corresponding to $R \geq 0.60$~m); see Visualization 2 for the corresponding video.
		Insets show the corresponding MCF geometry. 
	}
	\label{fig:Fig3}
\end{figure}

As a first demonstration we show how a focus at the distal end of the MCF is maintained independently of MCF bends.
To do so, we use the experimental setup detailed in the Supplemental Information to establish a focus at the output end of MCF2. In Fig.~\ref{fig:Fig3}(a) we show an image of the focus established for MCF2 held straight ($R$~=~$\infty$). Subsequently, and without changing the injection, we then displaced the distal end to change the radius of curvature of the MCF to $R$~=~0.06~m. The focus, as seen in Fig.~\ref{fig:Fig3}(b), retains its position to within one spot size. As a base of comparison, Figs.~\ref{fig:Fig3}(c), \ref{fig:Fig3}(d) show that the PSF of MCF0 is translated by a much larger amount for a miniscule conformational change (to $R$~=~0.6~m).
A side-by-side comparison is given by Visualization 1 (MCF2) and Visualization 2 (MCF0) which show how the PSF translated when the MCFs undergo the same conformational change. 
This showcases the dramatic reduction in extrinsic contribution in MCF2 to the level where MCF2 can be considered virtually conformationally invariant. 

%\subsection{Imaging} \label{sec:StaticImaging}
\begin{figure}[htbp]
	\centering
	\includegraphics[width = \textwidth]{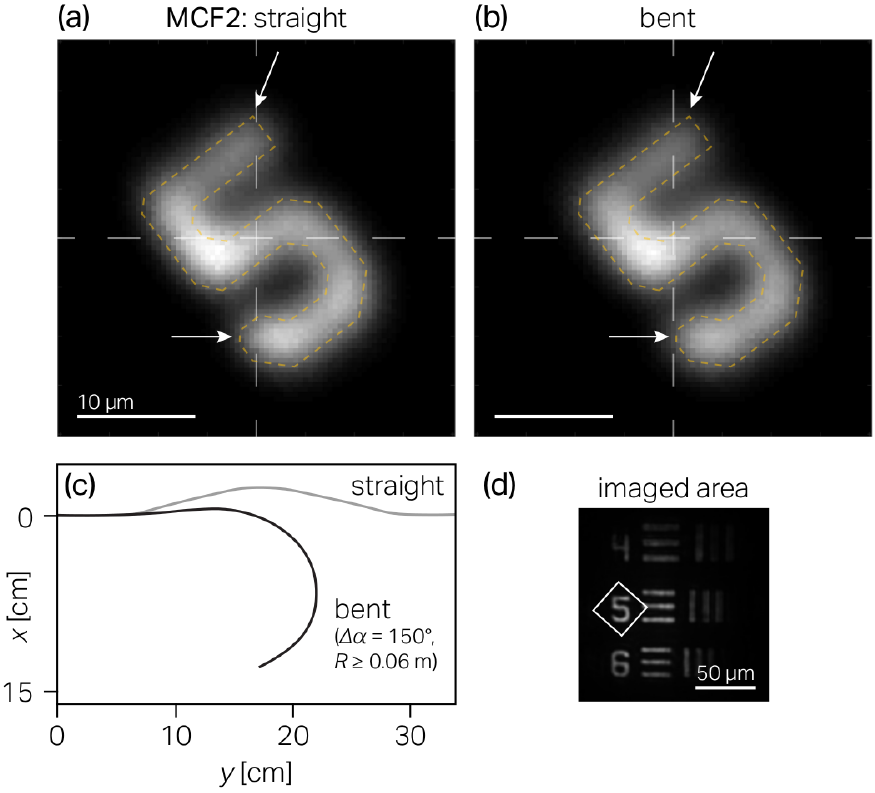}
	\caption{(a) Digital confocal TPEF image of a test target, obtained using the twisted MCF2 ($P$~=~8.2~mm) with $L=0.345$~m and $R=\infty$ and
		(b) with $R \geq 0.06$~m. See text for details.  %[Reference to video S1]
		(c) MCF2 experimental geometries for experiments (a,b). $\Delta \alpha$ is the angle between the input and output end of the MCF.
		(d) Imaged area mapped from a widefield target illumination.
	}
	\label{fig:Fig4}
\end{figure}
In Fig.~\ref{fig:Fig4} we show digital confocal images of a USAF target acquired through MCF2 in (Fig.~\ref{fig:Fig4}(a)) straight geometry ($R$~=~$\infty$); and (Fig.~\ref{fig:Fig4}(b)) bent geometry ($R$~=~0.06~m). A slight translation of the image can be observed which is reminiscent of the translation of the PSF from Fig.~\ref{fig:Fig3}(a) to Fig.~\ref{fig:Fig3}(b), and thus demonstrates virtually conformationally invariant lensless endoscopic imaging. 

\section{Discussion}
\label{sec:Discussion}
We note that the images presented here [Figs.~\ref{fig:Fig4}(a),\ref{fig:Fig4}(b)] are with one-photon contrast. The reason for this is the large native group delay that results from the twist [Fig.~\ref{fig:Fig2}(a)]. This native group delay is static and could be compensated with a static pre-compensation so, while not demonstrated here,  there are no conceptual difficulties in rendering the setup compatible with ultra-short laser excitation, and to access nonlinear imaging modalities. 

We note also that the imaging results presented here were done with forward-collection of signals, rather than with epi-collection through the MCF. The reason for this is the low fill factor of the cores, which results in poor epi-collection efficiency. However, it can be envisioned to increase the epi-collection efficiency of the twisted MCF by adding an external low-index cladding, like in e.g. Ref.~\cite{AndresenOE2013}.

In the above we have shown that the lensless endoscope based on twisted MCF with an integer number of twist periods is conformationally invariant as long as the bend radius of curvature remains constant. In practice there are two main ways that conditions can depart from this ideal condition: (i) Mismatch between length and twist period i.e. $L \ne k \cdot P, k \in N$; and (ii) non-constant bend radius of curvature i.e. $R$ varies along the MCF. Concerning (i) we present in the Supplemental Information a map of the extrinsic contribution to phase delay $\Delta[\Delta \phi^{(i)}]$ in the relevant parameters, permitting to identify the parametric sub-space where departure from conformational invariance does not exceed a certain threshold. Concerning (ii) it is not possible to offer insights from analytical considerations, instead an integral must be numerically evaluated, and in the Supplemental Information we calculate $\Delta[\Delta \phi^{(i)}]$ for realistic MCF shapes. We concede that conformational invariance naturally cannot be maintained over the entire parameter space, nevertheless we maintain that the twisted MCF drastically reduces the extrinsic contribution over the majority of fiber conformations likely to be found in practical settings.

Finally we remark that the shorter the twist period of a MCF, the more conformationally invariant it will be. In the limiting case of infinitesimal twist period, the MCF is conformationally invariant over the entire parameter space. In practice, this limit is unattainable because losses due to the curved path of the cores will set in at a certain twist period. In the Supplemental Information we detail numerical calculations which show that, for the MCF parameters used here, twist periods below 2~mm lead to losses above 1~dB/m.

%\section{Conclusion} \label{sec:Conclusion}
We have demonstrated a lensless endoscope based on a novel, twisted MCF at an operating point where imaging performance is unaffected by the conformation of the fiber. 
Our findings relax the constraints on fiber shape in lensless endoscopes and pave the way towards imaging tools that reap simultaneously the two principal benefits of using nothing but an optical fiber as imaging probe, namely, its small diameter and its flexibility.

\section{Methods}\label{sec:Methods}
\subsection{Fabrication of twisted MCF}
The MCF was made using the stack and draw method.  More precisely, a stack of 487 capillaries drawn from a commercial pure silica tube (Heraeus F300) were first assembled inside a silica tube of $\sim$50~mm outside diameter. Then 487 rods drawn from a commercial graded-index preform ($\Delta n$~=~30 $\cdot$ 10$^{-3}$, Prysmian) were inserted into each of these capillaries. This “sleeving” approach was used in order to guarantee no coupling between adjacent cores of the final fiber by increasing sufficiently the distance between these cores. This stack was drawn into canes of about 5~mm. During this draw, a vacuum was applied between the capillaries and inside them in order to get solid canes free of bubbles that could appear at the different silica interfaces present in the stack. Finally one of these canes was drawn into a fiber of $\sim$450~$\mu$m diameter at low drawing speed (2~m/min), the twist being obtained by rotating the cane at the top of the fiber drawing tower (60~rpm for MCF1 and 250~rpm at MCF2).
\subsection{Experimental methods}
The experimental setup (Fig. \ref{fig:FigA-setup}, Supplemental Information) was designed to measure the MCF properties using two modalities: ultrashort (fs) pulses or continuous wave (CW) in order to perform group delay and imaging measurements, respectively.
The first modality was used to perform the phase-stepping spectral interferometry to measure group delays of laser pulses transmitted through different MCF cores (with respect to a reference core, usually the central core), as described in \cite{AndresenJOSAB2015}.
The second modality is used to measure the PSF and imaging performance during MCF conformational changes. 

For phase-stepping spectral interferometry (Fig.~\ref{fig:FigA-setup}, Supplemental Information) the laser beam (Amplitude Syst\`{e}mes t-Pulse, central wavelength 1030 nm, pulse length 170 fs, repetition rate 50 MHz) is expanded with a telescope (lenses L1 and L2) to overfill the clear aperture of a spatial light modulator (SLM, Hamamatsu LCoS-SLM X10468-07). 
The SLM is used to segment and shape the wavefront into beamlets prior to their injection into the cores of the MCF proximal facet. For group delay measurements only two cores are injected into (the central, reference, core and the core $i$) resulting in fringes at the MCF output far field. For PSF measurements all cores are injected into in order to produce a focus at the MCF output. 
The necessary demagnification of the beamlets to fit the core diameters is achieved via a lens L3 and a microscope objective MO1 (Olympus Plan N, 20x NA 0.40).
Light, transmitted through the cores, is collected at the fiber distal end using a second microscope objective MO2 (Nikon Plan, 10x NA 0.25) and passes through a linear polarizer LP (Thorlabs LPNIR100) to maximize the contrast of the interference fringes \cite{SivankuttyOL2016}.
The image of the distal MCF facet or its near field is monitored with CCD1 (FLIR FL3-U3-32S2M-CS) in order to follow any changes of the transmitted power or the generated interference pattern while the fiber is bent.
The MCF output far field is coupled into a multimode fiber (core diameter of 62.5 $\mu$m, not shown in Fig.~\ref{fig:FigA-setup}, Supplemental Information), linked to an optical spectrum analyzer OSA (Yokogawa AQ-6315A), or imaged entirely onto a camera CCD2 (Thorlabs DCU223M) to record the point spread function (PSF) stability in during MCF bending.
Magnification of the objective MO2 and lenses L4, L5 is chosen so that only restricted area of the far field pattern is selected with the MMF probe (fulfilling therefore $k_{x} D > 2$, where $k_{x}$ is related to the interference fringe spatial frequency and $D$ is diameter of the MMF core collecting the light).
The distal end of the MCF is kept fixed on a portable unit that can freely rotate whilst the MCF is bent. 
The fiber conformation is monitored using a portable camera (not shown). Finally galvanometric scan mirrors conjugated to the SLM active area (telescope L1 and L2) are used to scan the focused distal PSF across the sample for imaging. The galvanometric mirrors impose controllable phase tilts on the input wavefront that are translated to the output distal MCF facet owing to the diagonal MCF transmission matrix. 

\section*{Funding Information}
Institut Fresnel:
Agence Nationale de la Recherche ANR-14-CE17-0004-01; ANR-10-INSB-04-01; ANR-11-INSB-0006; ANR-11-IDEX-0001-02.
Institut National de la Sant\'{e} et de la Recherche M\'{e}dicale PC201505, 18CP128-00.
EU Eurostar 12583/19/Q DEEPBRAIN.
SATT Sud-Est GDC Lensless endoscope.
CNRS-Weizmann ImagiNano European Associated Laboratory. \\
Research reported in this publication was supported by the National Eye Institute of the National Institutes of Health under Award Number R21EY029406. The content is soleley the responsibility of the authors and does not necessarily represent the official views of the National Instituted of Health.

PhLAM Laboratory: Agence Nationale de la Recherche
LABEX CEMPI (ANR-11-LABX-0007);
Equipex Flux (ANR-11-EQPX-0017).
Ministry of Higher Education and Research, Hauts de France council and European Regional Development Fund (ERDF) through the Contrat de Projets Etat-Region (CPER Photonics for Society P4S).\\

\section*{Acknowledgments}

\section*{Supplemental Documents}
\noindent Supporting Material available: Visualization 1, Visualization 2, and Supplemental Information.

% Bibliography

\clearpage

\section*{Supplemental Information}

\section{Experimental setup} \label{sec:ExperimentalSetup}
\begin{figure*}[htbp]
	\centering
	\includegraphics[width = \textwidth]{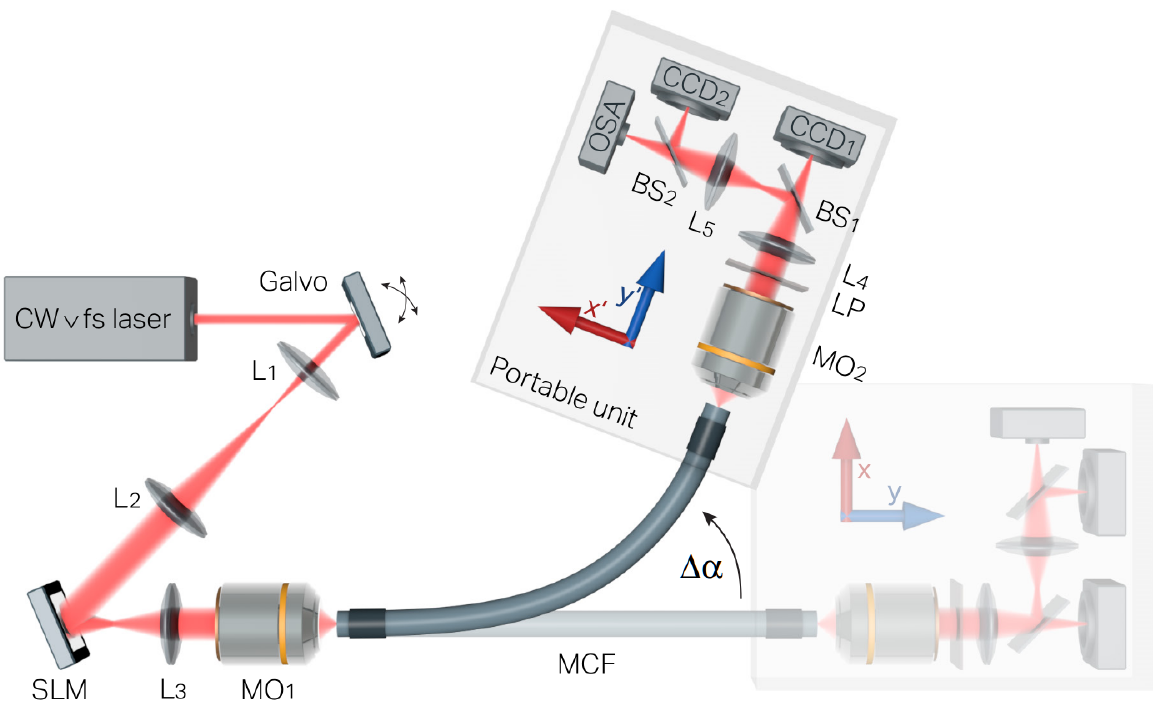}
	\caption{Schematic of the experimental setup used in the experiments.
	CW, continuous wave; fs, femtosecond; galvo, galvanometric mirror; L, lens; SLM, spatial light modulator; MO, microscope objective; MCF, multicore fiber; LP, linear polarizer; BS, beam splitter; CCD, charge coupled device; OSA, optical spectrum analyzer. 
	The MCF (in gray) is clamped at both ends along approximately 2 cm long sections (dark gray).
	}
	\label{fig:FigA-setup}
\end{figure*}

The experimental setup (Fig. \ref{fig:FigA-setup}) was designed to measure the MCF properties using two modalities: ultrashort (fs) pulses or continuous wave (CW) in order to perform group delay and imaging measurements, respectively.
The first modality was used to perform the phase-stepping spectral interferometry to measure group delays of laser pulses transmitted through different MCF cores (with respect to a reference core, usually the central core), as described in \cite{AndresenJOSAB2015}.
The second modality is used to measure the PSF and imaging performance during MCF conformational changes. 

For phase-stepping spectral interferometry (Fig.~\ref{fig:FigA-setup}) the laser beam (Amplitude Syst\`{e}mes t-Pulse, central wavelength 1030 nm, pulse length 170 fs, repetition rate 50 MHz) is expanded with a telescope (lenses L1 and L2) to overfill the clear aperture of a spatial light modulator (SLM, Hamamatsu LCoS-SLM X10468-07). 
The SLM is used to segment and shape the wavefront into beamlets prior to their injection into the cores of the MCF proximal facet. For group delay measurements only two cores are injected into (the central, reference, core and the core $i$) resulting in fringes at the MCF output far field. For PSF measurements all cores are injected into in order to produce a focus at the MCF output. 
The necessary demagnification of the beamlets to fit the core diameters is achieved via a lens L3 and a microscope objective MO1 (Olympus Plan N, 20x NA 0.40).
Light, transmitted through the cores, is collected at the fiber distal end using a second microscope objective MO2 (Nikon Plan, 10x NA 0.25) and passes through a linear polarizer LP (Thorlabs LPNIR100) to maximize the contrast of the interference fringes \cite{SivankuttyOL2016}.
The image of the distal MCF facet or its near field is monitored with CCD1 (FLIR FL3-U3-32S2M-CS) in order to follow any changes of the transmitted power or the generated interference pattern while the fiber is bent.
The MCF output far field is coupled into a multimode fiber (core diameter of 62.5 $\mu$m, not shown in Fig.~\ref{fig:FigA-setup}), linked to an optical spectrum analyzer OSA (Yokogawa AQ-6315A), or imaged entirely onto a camera CCD2 (Thorlabs DCU223M) to record the point spread function (PSF) stability in during MCF bending.
Magnification of the objective MO2 and lenses L4, L5 is chosen so that only restricted area of the far field pattern is selected with the MMF probe (fulfilling therefore $k_{x} D > 2$, where $k_{x}$ is related to the interference fringe spatial frequency and $D$ is diameter of the MMF core collecting the light).
The distal end of the MCF is kept fixed on a portable unit that can freely rotate whilst the MCF is bent. 
The fiber conformation is monitored using a portable camera (not shown). Finally galvanometric scan mirrors conjugated to the SLM active area (telescope L1 and L2) are used to scan the focused distal PSF across the sample for imaging. The galvanometric mirrors impose controllable phase tilts on the input wavefront that are translated to the output distal MCF facet owing to the diagonal MCF transmission matrix.

\section{Bending-induced inter-core group delays for twisted and non-twisted fiber} \label{sec:SpatialGDD}
\begin{figure}[htbp]
	\centering
	\includegraphics[width = \textwidth]{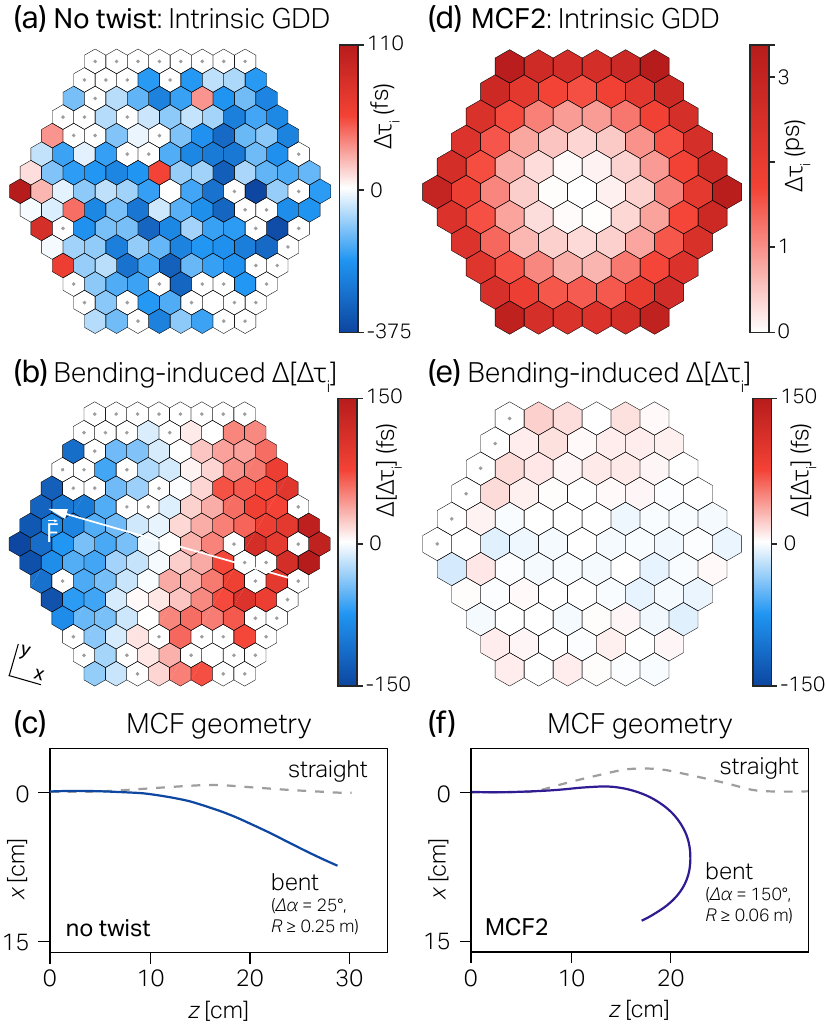}
	\caption{Measured group delays $\Delta\tau_i$ as a function of transverse core position for (a) non-twisted MCF0 and (b) twisted MCF2 ($P$~=~8.2~mm). 
		(b,c) Change in the group delay $\Delta[\Delta\tau_i]$ for MCF0 experiencing a $\Delta \alpha$~=~25$^\circ$ angle. (e,f)Change in $\Delta[\Delta\tau_i]$ for MCF2 experiencing $\Delta \alpha$~=~150$^\circ$. 
		White arrow in (b) indicates the direction of applied force during the bending.
		Maximum absolute values $\Delta[\Delta\tau_i]$ in (e) are less than 20~fs, the resolution of the phase-stepping spectral interferometry technique.
		Experimental data for the non-twisted MCF0 was taken from \cite{TsvirkunOE2017}.
	}
	\label{fig:FigB-GDD}
\end{figure}

Figure~\ref{fig:FigB-GDD} provides additional experimental data for the native and extrinsic contribution to group delay in the cores, presented as a function of the transverse core coordinates.
In the case of the non-twisted fiber MCF0 held straight [Fig.~\ref{fig:FigB-GDD}(a)] we observe a random distribution of group delays with standard deviation [XXX - VT please specify the standard deviation] that is related to intrinsic imperfections \cite{TsvirkunOE2017, AndresenJOSAB2015}. 
For a relatively small angle ($\Delta\alpha = 25 ^{\circ}$) 	[Fig.~\ref{fig:FigB-GDD}(c)] we observe a linear dependence of the bending-induced group delays on the core position with respect to the bending axis, as detailed in \cite{TsvirkunOE2017}. 
If MCF0 were employed with ultrashort pulses, increasing $\Delta\alpha$ will result in a PSF degradation associated with a non perfect temporal overlap of the pulses arriving at the focal plane of a lensless endoscope. 

The twisted MCF2, on the other hand, exhibits important native group delays when held straight [Fig.~\ref{fig:FigB-GDD}(d)] due to the increased physical lengths of the peripheral cores (Eq.~\ref{eq:Deltataui}).
Extrinsic contribution do to conformational change is, however, negligible in this fiber [Fig.~\ref{fig:FigB-GDD}(e)] and did not exceed $\pm20$~fs for a bending angle of $\Delta\alpha = 150 ^{\circ}$, which is close to the resolution limit of the used phase-stepping spectral inferferometry technique. 
Indeed Figs.~\ref{fig:Fig3}(a),\ref{fig:Fig3}(b) (main text) showed that the extrinsic contribution to phase delay is less than 2$\pi$, this strongly suggests that extrinsic contribution to group delay remains less than an optical cycle or 3.3~fs. 

\section{Fabrication of twisted MCF} \label{sec:Fabrication}
The MCF was made using the stack and draw method.  More precisely, a stack of 487 capillaries drawn from a commercial pure silica tube (Heraeus F300) were first assembled inside a silica tube of $\sim$50~mm outside diameter. Then 487 rods drawn from a commercial graded-index preform ($\Delta n$~=~30 $\cdot$ 10$^{-3}$, Prysmian) were inserted into each of these capillaries. This “sleeving” approach was used in order to guarantee no coupling between adjacent cores of the final fiber by increasing sufficiently the distance between these cores. This stack was drawn into canes of about 5~mm. During this draw, a vacuum was applied between the capillaries and inside them in order to get solid canes free of bubbles that could appear at the different silica interfaces present in the stack. Finally one of these canes was drawn into a fiber of $\sim$450~$\mu$m diameter at low drawing speed (2~m/min), the twist being obtained by rotating the cane at the top of the fiber drawing tower (60~rpm for MCF1 and 250~rpm at MCF2).

\section{Analytical model of twisted MCF}
\label{sec:AnalyticalModel}
The analytical model is equivalent to the first order of perturbation for which the field distribution is not affected by the twist. Since the refractive index contrast of the fiber is very low, we can consider that the effective index is equal to the group index : with the parameters given in part \ref{sec:TwistedMCF}A, $n_{eff}=1.4626$ and $n_g=1.4909$ for the central core at 1 $\mu$m wavelength.

\subsection{Preamble}
Using the coordinate system and definitions of Fig.~\ref{fig:Fig1} we write the following parametric equations for the coordinates of core $i$ which is at the distance $d^{(i)}$ from the center of the fiber and which makes a $\xi^{(i)}$ angle with the normal vector directed outside the curve described by the fiber:
\begin{eqnarray}
	x^{(i)}(\psi) &=& [R + d^{(i)} \, \mathrm{cos} ( \frac{2\pi}{P} R \psi + \xi^{(i)})] \, \mathrm{cos}\psi \\
	y^{(i)}(\psi) &=& [R + d^{(i)} \, \mathrm{cos} ( \frac{2\pi}{P} R \psi + \xi^{(i)})] \, \mathrm{sin}\psi \\
	z^{(i)}(\psi) &=& d^{(i)} \, \mathrm{sin}(\frac{2\pi}{P}R \psi + \xi^{(i)} )
\end{eqnarray}
from which we can find the physical length of core $i$ by the line integral
\begin{eqnarray}
	L^{(i)} &=& \int_{0}^{\frac{L}{R}} \sqrt{\Big ( \frac{\mathrm{d}x}{\mathrm{d}\psi} \Big )^{2} + \Big ( \frac{\mathrm{d}y}{\mathrm{d}\psi} \Big )^{2} + \Big ( \frac{\mathrm{d}z}{\mathrm{d}\psi} \Big )^{2} } \mathrm{d}\psi  \\
  	&=& \int_{0}^{\frac{L}{R}} \sqrt{(\frac{2 \pi}{P} d^{(i)} R)^{2} + \{ R + d^{(i)} \mathrm{cos} [\frac{2 \pi}{P} R \psi + \xi^{(i)}) ] \}^2} \mathrm{d}\psi
\end{eqnarray}

\subsection{Case: Straight MCF}
In the special case of an unbent MCF $R$~=~$\infty$ and the expression reduces to
\begin{equation}
L^{(i)} = L \, \frac{\sqrt{(2 \pi d^{(i)})^{2} + P^2}}{P}
\end{equation}
which is how Eq.~\ref{eq:Li_straight} comes about.
The optical path length of core $i$ is
\begin{equation}
\mathrm{OPL}_{\mathrm{straight}}^{(i)} = n_{\mathrm{eff}}^{(i)} L^{(i)} = n_{\mathrm{eff}}^{(0)} L^{(i)}.
\end{equation}
We note that we can take $n_{\mathrm{eff}}^{(i)}$~=~$n_{\mathrm{eff}}^{(0)}$ because the twist of the MCF is introduced during the fiber drawing process, so no additional stress accompanies the twist. This is different from the case where twist is introduced by mechanical twisting of the MCF in which case the photoelastic effect must be taken into account.
The group delay experienced by a pulse of light travelling in core $i$ is then
\begin{equation}
\tau^{(i)} = \frac{\mathrm{OPL}^{(i)}}{c}
\end{equation}
or, relative to the center core with $i$~=~0:
\begin{eqnarray}
\Delta \tau^{(i)} &=& \tau^{(i)} - \tau^{(0)} \\
  &=& \frac{n_{\mathrm{eff}}^{(0)}}{c}(L^{(i)} - L).
\end{eqnarray}
Which is how Eq.~\ref{eq:Deltataui} comes about. \\
And similarly, the phase delay
\begin{equation}
\Delta \phi^{(i)} = \phi^{(i)} - \phi^{(0)} = \frac{2 \pi}{\lambda} n_{\mathrm{eff}}^{(0)}  (L^{(i)} - L)
\end{equation}
Which is how Eq.~\ref{eq:Deltaphii} comes about. \\
\subsection{Bent MCF}
\begin{figure}[htbp]
	\centering
	\includegraphics[width=\textwidth]{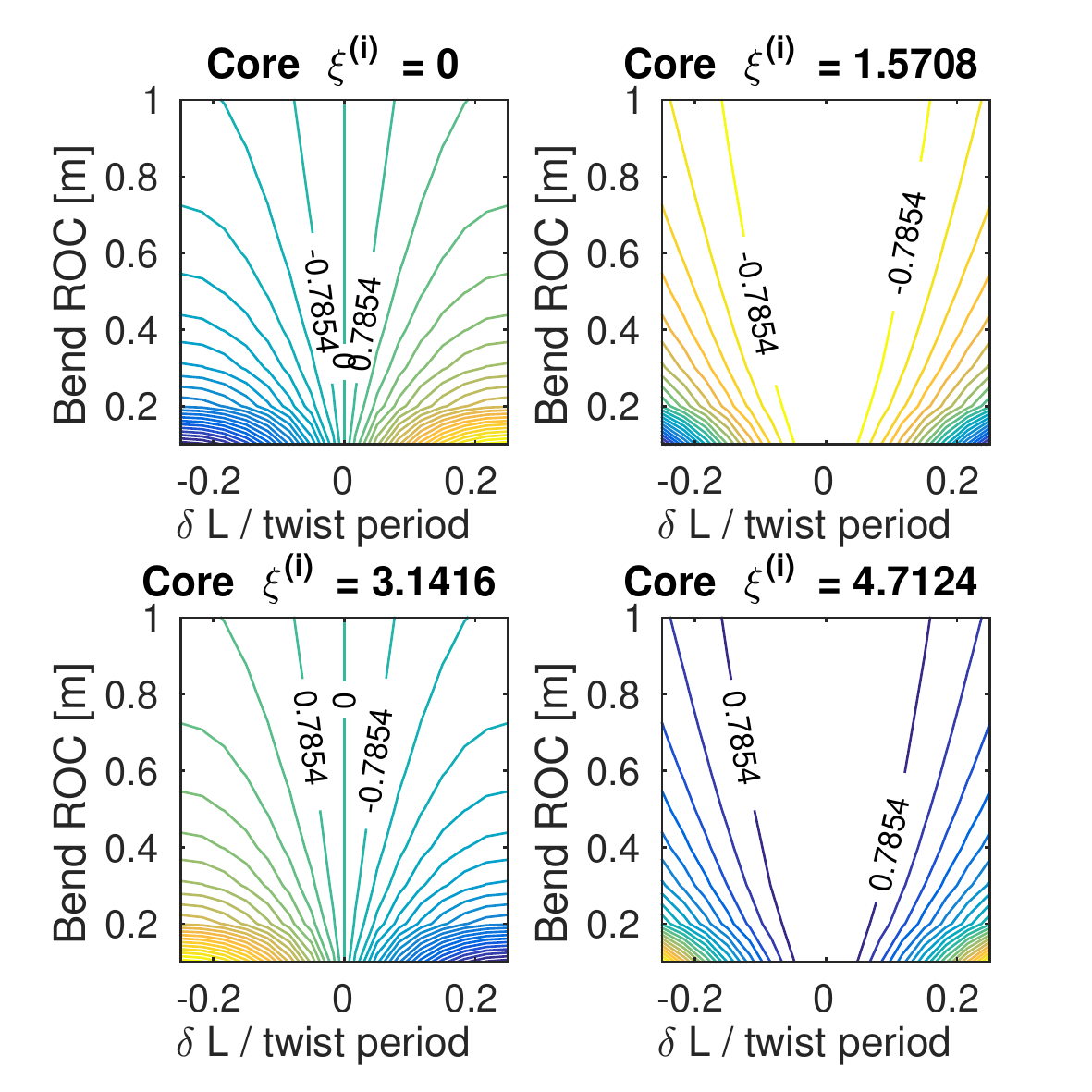}
	\caption{
		Contour plots of $\delta \phi^{(i)}$ for the extreme core with $d^{(i)}$ = 192~$\mu$m and varying $\xi^{(i)}$. Bend ROC corresponds to $R$ in Fig. \ref{fig:Fig1}.
	}
	\label{fig:dphi}
\end{figure}

In the general case $R \ne \infty $, $L^{(i)}$ we must take into account the photoelastic effect of silica. Indeed, the refractive index of the core $i$ is modified as:
\begin{equation}
n_{\mathrm{bent}}^{(i)} = n_{\mathrm{core}} \Bigg ( 1 - p_{e}\frac{d^{(i)} \mathrm{cos} (\frac{2\pi}{P}R\psi) }{R} \Bigg )
\end{equation}
where $p_e\simeq 0.22$ for fused silica is a coefficient depending on Poisson's ratio and components of the photoelastic tensor \cite{SchermerJQE2007}.\\ 
If the radius of the curvature is much smaller than the fiber diameter, we can use the first order theory of perturbation and assume that the electric field $E$ is not affected by the bending and remains confined in the core. In the scalar approximation \cite{SnyderLove}:
\begin{equation}
\delta \left(n_{\mathrm{eff}}^{(i)}\right)^2=\frac{\int \int \delta n^2 \vert E \vert^2 \mathrm{d}S}{\int \int \vert E\vert^2 \mathrm{d}S}\simeq \delta \left(n_{\mathrm{core}}^{(i)}\right)^2
\end{equation}
Since $n_{\mathrm{eff}}\simeq n_{\mathrm{core}}$ in the scalar approximation, the effective index is modified as:
\begin{equation}
n_{\mathrm{eff}}^{(i)} = n_{\mathrm{eff}}^{(0)} \Bigg ( 1 - p_{e}\frac{d^{(i)} \mathrm{cos} (\frac{2\pi}{P}R\psi) }{R} \Bigg )
\end{equation}
The group index is modified in the same manner since \cite{SnyderLove}:
\begin{equation}
n_{g} = \frac{\int \int |E|^2 (n^2 + n_{\mathrm{eff}}^2 - \lambda \frac{\mathrm{d}n^2}{\mathrm{d}\lambda}) \mathrm{d}S}{2 n_{\mathrm{eff}} \int \int |E|^2 \mathrm{d} S}
\end{equation}
hence
\begin{equation}
\delta n_g = (2 - \frac{n_g}{n_{\mathrm{eff}}}) \delta n_{\mathrm{eff}}\simeq \delta n_{\mathrm{eff}}
\end{equation}
As a result, the optical path length of core $i$ of the bent, twisted fiber is
\begin{eqnarray}
\mathrm{OPL}_{\mathrm{bent}}^{(i)} &=& \int_{0}^{\frac{L}{R}} n_{\mathrm{eff}}^{(0)} \Bigg ( 1 - p_{e}\frac{d^{(i)} \mathrm{cos} (\frac{2\pi}{P}R\psi) }{R} \Bigg ) \\
&& \times \sqrt{(\frac{2 \pi}{P} d^{(i)} R)^{2} + \{ R + d^{(i)} \mathrm{cos} [\frac{2 \pi}{P} R \psi + \xi^{(i)}) ] \}^2} \mathrm{d}\psi \nonumber
\end{eqnarray}
We know that $d^{(i)}/R \ll 1$, so we can get some initial insights from a linear expansion,
\begin{eqnarray}
&&\frac{\mathrm{OPL}_{\mathrm{bent}}^{(i)}}{n_{\mathrm{eff}}^{(0)}}   = L\sqrt{1+\left(\frac{2\pi d^{(i)}}{P}\right)^2} + \frac{d^{(i)}}{R}\frac{P}{2\pi} \Bigg( \frac{1}{\sqrt{1+(\frac{2\pi d^{(i)}}{P})^2}} - \nonumber \\
&&p_{e}\sqrt{1+(\frac{2\pi d^{(i)}}{P})^2}  \Bigg )
[ \mathrm{sin}(\frac{2\pi L}{P} + \xi^{(i)}) - \mathrm{sin}(\xi^{(i)}) ]
\end{eqnarray}
From which it can be seen that when $L$ = $k \cdot P$, $k \in N$, OPL$^{(i)}$ does not change with $R$.
This is the basis of the conformationally invariant lensless endoscope as demonstrated in the main text.
In practice, however, it is not possible to attain this condition with arbitrary precision.
In order to remain within the regime of conformational invariance, the following criterion on the error phase $\delta \phi^{(i)}$ 
\begin{equation}
\label{eq:errorphase}
\delta \phi^{(i)} = \frac{2\pi}{\lambda}(\mathrm{OPL}_{\mathrm{bent}}^{(i)} - \mathrm{OPL}_{\mathrm{straight}}^{(i)})
\end{equation}
should respect the criterion $\delta \phi^{(i)} < \pi/2, \forall i$. Eq.~\ref{eq:errorphase} can in principle be evaluated to arbitrary precision as a function of all involved parameters ($P$, $\xi^{(i)}$, $L^{(i)}$, $R$, $d^{(i)}$). 
Here, we evaluate $\delta \phi^{(i)}$ for an excursion $\delta L$ in a range around the optimal value of $L$ and bend radius of curvature $R$, mapped in Fig.~\ref{fig:dphi} for the extreme core with $d^{(i)}$~=~12$\cdot$16~=~192~$\mu$m which experiences the largest error phase of all the cores.

\subsection{Non-constant bend radius of curvature}
\begin{figure}[htbp]
	\centering
	\includegraphics[width = \textwidth]{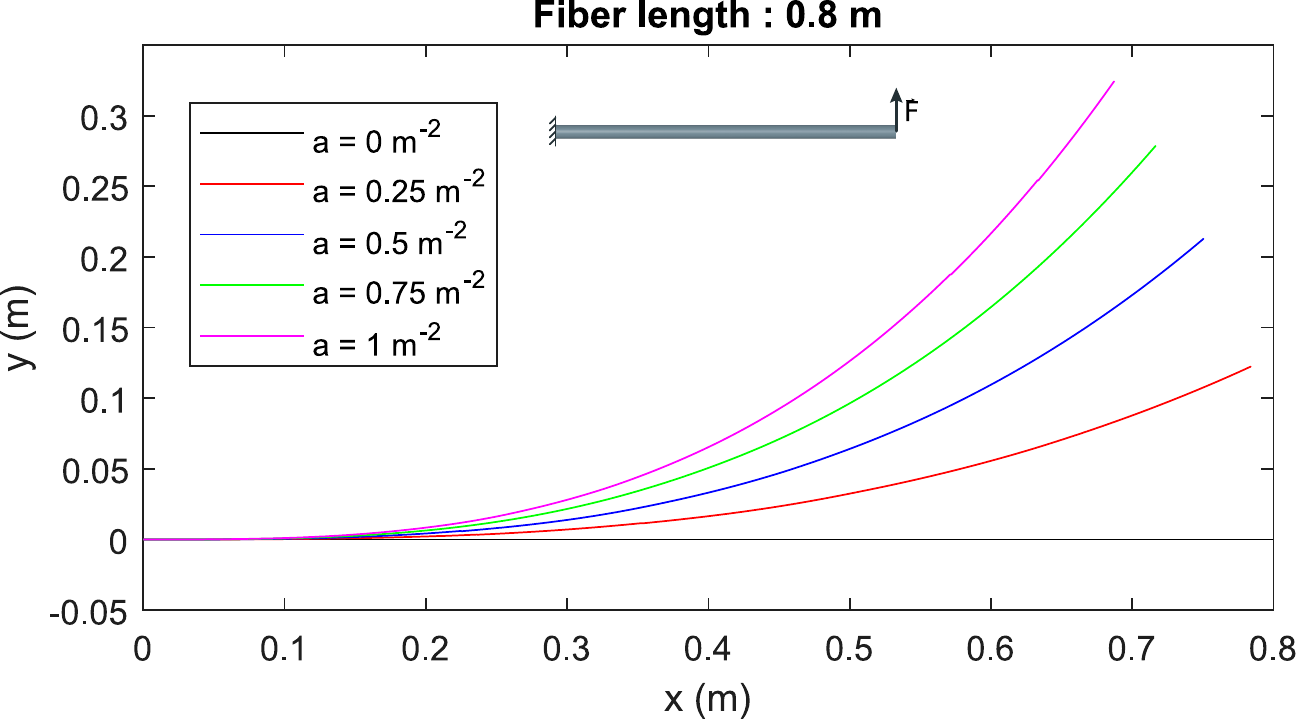}
	\caption{Different shapes of the fiber when applying a normal force on one extremity.
	}
	\label{fig:forme}
\end{figure}

All the results obtained with a constant radius can be easily extended by replacing the constant radius $R$ with the radius of curvature of the curve described by the fiber. No general results can be extracted from the analytical formulae and, as an example, we will study the case of a fiber maintained fixed at one extremity and whose other end is subjected to a normal force. It is well-known that, in this case, the curve of the fiber is described by a cubic function $y=ax^3$ ($a$ depends on the force, the Young modulus and the second moment of area of the cross section of the fiber). Some different bending configuration are depicted in Fig \ref{fig:forme} for different value of $a$ and a constant fiber length of 0.8~m.

\begin{figure}[htbp]
	\centering
	\includegraphics[width = \textwidth]{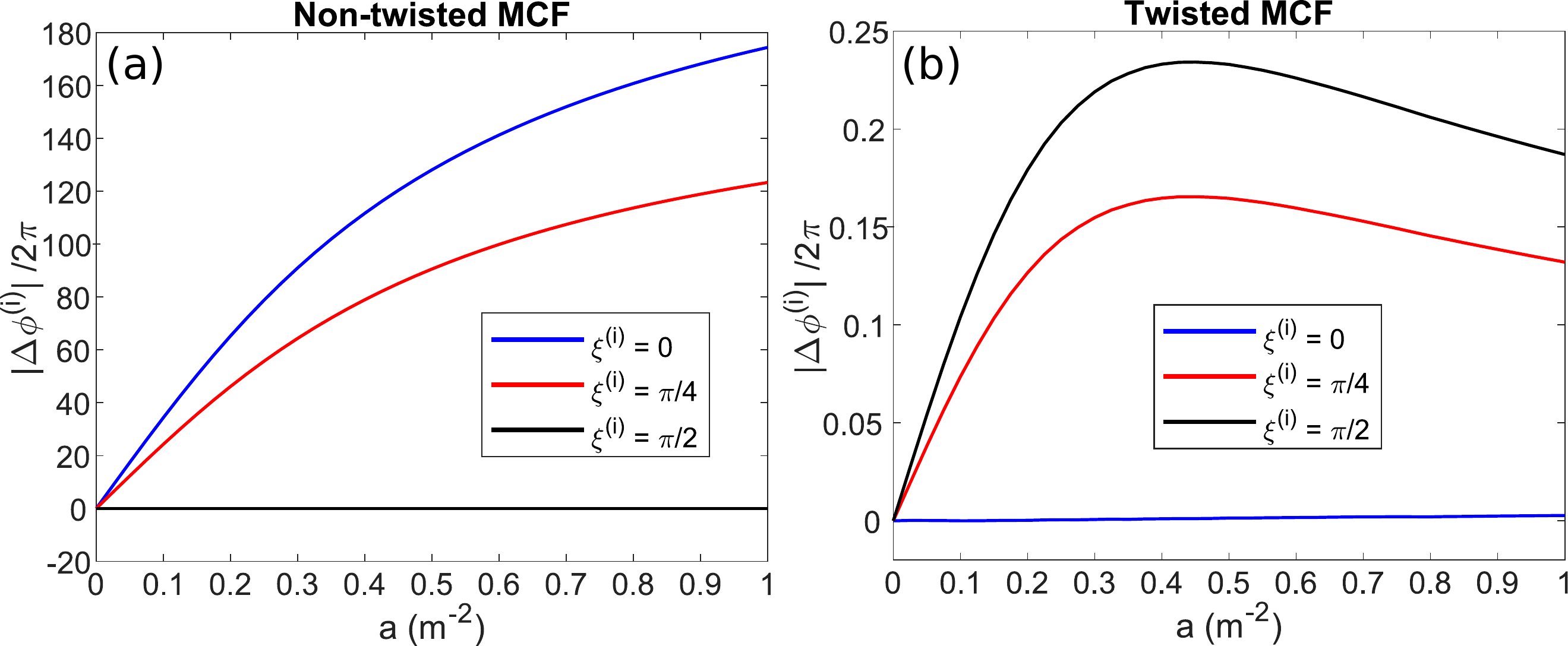}
	\caption{Variation of $\Delta \phi^{(i)}$ with the coefficient $a$ for (a) non-twisted MCF and (b) 8 mm-period twisted MCF if $d^{(i)}= 160 $~$\mu$m.}
	\label{fig:deltaphi}
\end{figure}

The variation of phase $\Delta \phi^{(i)}$ (defined in Eq. 3) between the core $i$ and the central core was numerically computed and plotted if $d^{(i)}= 160 $~$\mu$m for a non-twisted fiber (Fig \ref{fig:deltaphi}a) and for a twisted fiber with a helical period of 8~mm (Fig \ref{fig:deltaphi}b). For the non-twisted MCF, we have already demonstrated that \cite{TsvirkunOE2017}:
\begin{equation}
\frac{\vert \Delta \phi^{(i)}\vert}{2\pi}\simeq (1-p_e)\frac{d^{(i)}\vert \cos(\xi^{(i)})\vert}{\lambda} \Delta \alpha
\end{equation}
where $\Delta \alpha$ is the angular increase along the curve formed by the fiber.\\
This gives a phase difference of approximately $360\pi$ between the core $i$ and the central core if $a=1$~m$^{-2}$ and $\xi^{(i)}=0$ while the maximum phase difference in the case of the twisted MCF is only $0.47\pi$. This clearly demonstrates the bending resilience of the twisted MCF.

\section{FEM model of twisted MCF} \label{sec:NumericalModel}
To determine the properties of the modes with the finite elements method (FEM), we used method describe by Nicolet \textit{et al.} \cite{Nicolet2007}. This method uses the transformation optics formalism to obtain an equivalent translation invariant problem for which the permittivity and the permeability are anisotropic. The effective indices calculated with this method must be modified according to the mode azimutal number to obtain the real effective indices \cite{NapiorkowskiOE2014}. Note than this method is not applicable to a bent twisted fiber since the problem can not be reduced to a translation invariant problem in this case. However, this allows to determine mode losses due to the twist which are neglected in the analytical model. Note that the effective index and the group index calculated by this method are defined as: 
\begin{eqnarray}
n_{\mathrm{eff}}^{(i)} &=& \frac{\lambda}{2\pi}\frac{\phi^{(i)}}{L} \\
n_{\mathrm{g}}^{(i)} &=& \frac{c\tau^{(i)}}{L}
\end{eqnarray}

\begin{figure}[htbp]
\includegraphics[width=0.45\linewidth]{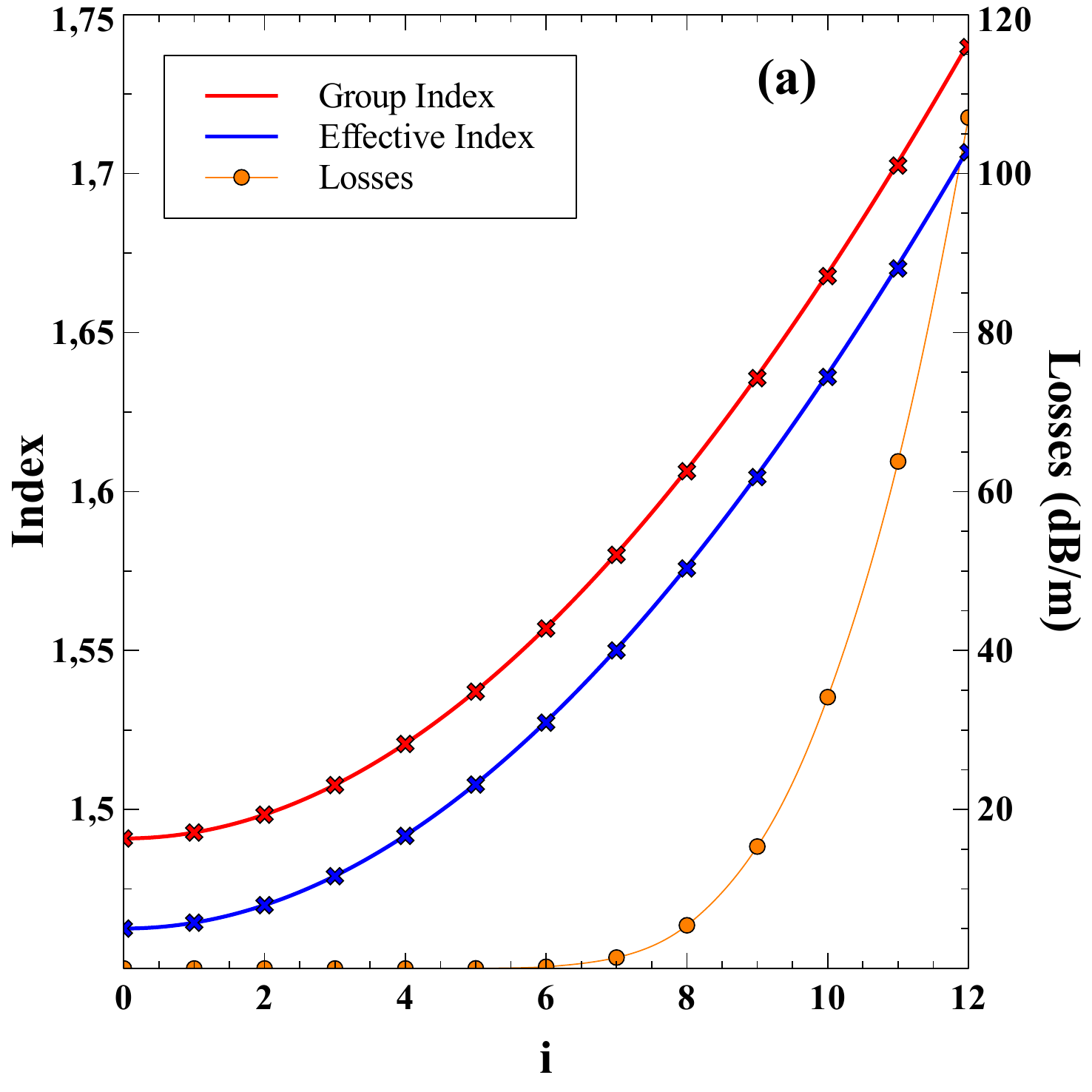}\includegraphics[width=0.45\linewidth]{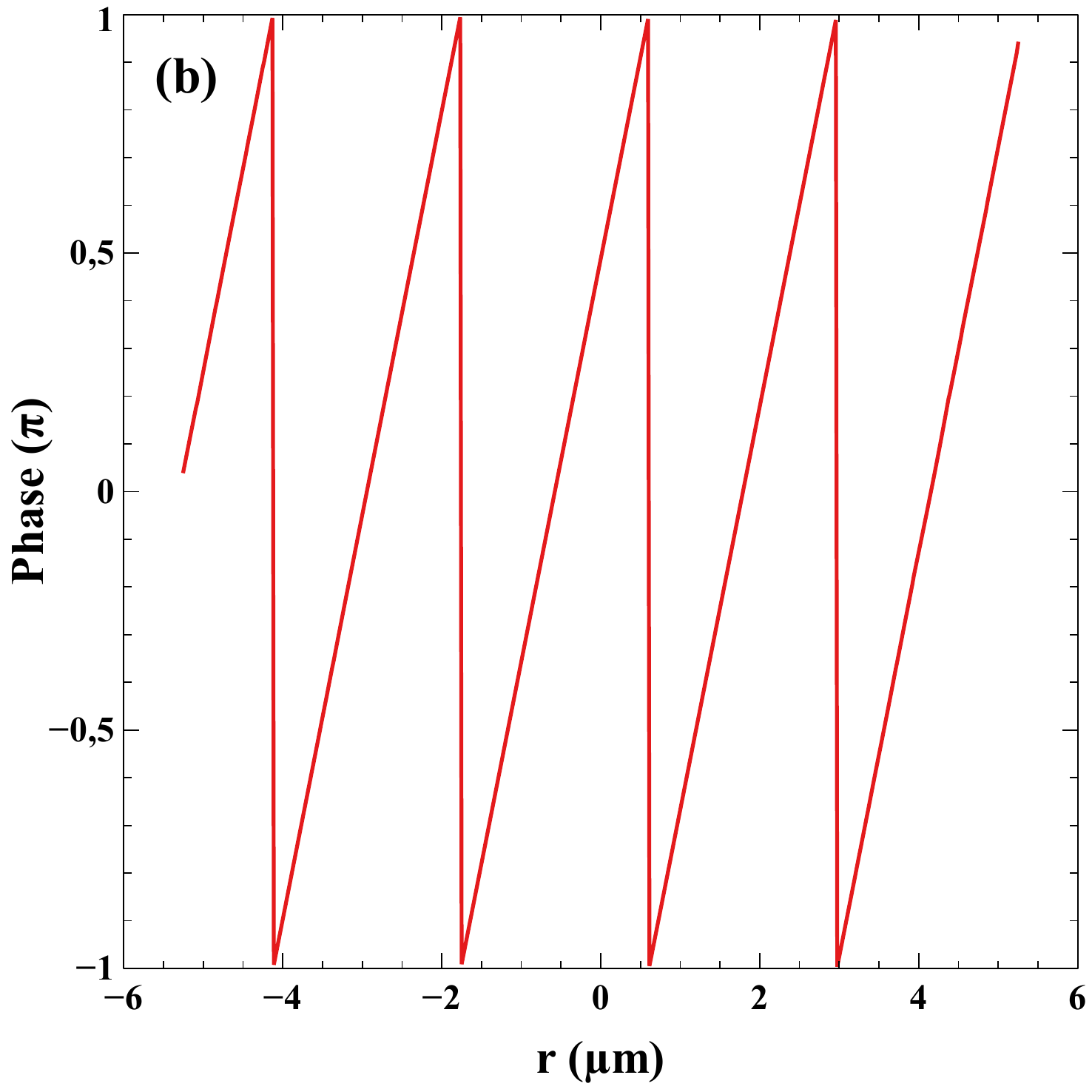}
\caption{\label{fig:comsol}
Numerical simulation. 
(a) Effective index (red), group index (blue) and losses (orange) of a twisted fiber whose parameters are described in section 2 with a helical period $P=2$~mm. Losses were computed with the FEM method (Comsol Multiphysics). Blue and red lines were obtained with Eq. 9 while blue and red crosses were obtained with the FEM method.(b) Phase of the electric field computed with the FEM method with $d^{(i)}/\Lambda=6$.}
\end{figure}

In Fig. \ref{fig:comsol}a, values of effective index and group index obtained with Eq.~9 (blue and red lines) are compared with those obtained with the FEM method (blue and red crosses). Losses of the fundamental mode computed with the FEM method are plotted in orange (right axis). We can verify that the approximations used to obtained Eq. 9 are valid with the parameters used in this article even when losses are very high ($>$~1dB/m). Note that losses are are lower than 1 dB/km when the twist period equals 8 mm.\\
The evolution of the phase of the electric field with the distance is plotted in Fig. \ref{fig:comsol}b. The linear evolution is due to the beam angle with the MCF axis : according to Eq. 5 when $\frac{2 \pi d^{(i)}}{P} \ll 1 $, the slope is 
\begin{equation}
\frac{d\phi_E}{dr}=n_{\mathrm{eff}}^{(0)} \frac{2 \pi d^{(i)}}{P}\frac{2\pi}{\lambda}
\end{equation}


\begin{thebibliography}{10}

\bibitem{ZivCurrOpinNeurobiol2015}
Y.~Ziv and K.~K. Ghosh, ``Miniature microscopes for large-scale imaging of
  neuronal activity in freely behaving rodents,'' {\em Curr. Opin. Neurobiol.},
  vol.~32, pp.~141--147, 2015.

\bibitem{LombardiniLSA2018}
A.~Lombardini, V.~Mytskaniuk, S.~Sivankutty, E.~Andresen, X.~Chen, J.~Wenger,
  M.~Fabert, N.~Joly, F.~Louradour, A.~Kudlinski, and H.~Rigneault,
  ``High-resolution multimodal flexible coherent raman endoscope,'' {\em Light:
  Science \& Applications}, vol.~7, p.~10, 2018.

\bibitem{ZongNatMeth2017}
W.~Zong, R.~Wu, M.~Li, Y.~Hu, Y.~Li, J.~Li, H.~Rong, H.~Wu, Y.~Xu, Y.~Lu,
  H.~Jia, M.~Fan, Z.~Zhou, Y.~Zhang, A.~Zang, L.~Chen, and H.~Cheng, ``Fast
  high-resolution miniature two-photon microscopy for brain imaging in freely
  behaving mice,'' {\em Nat. Meth.}, vol.~14, no.~7, pp.~713--722, 2017.

\bibitem{FlusbergNatMeth2005}
B.~A. Flusberg, E.~D. Cocker, W.~Piyawattanametha, J.~C. Jung, E.~Cheung,
  L.~M., and M.~Schnitzer, ``Fiber-optic fluorescence imaging,'' {\em Nat.
  Methods}, vol.~2, no.~12, pp.~941--950, 2005.

\bibitem{ThompsonOL2011}
A.~J. Thompson, C.~Paterson, M.~A.~A. Neil, C.~Dunsby, and P.~M.~W. French,
  ``Adaptive phase compensation for ultracompact laser scanning
  endomicroscopy,'' {\em Opt. Lett.}, vol.~36, no.~9, pp.~1707--1709, 2011.

\bibitem{CizmarOE2011}
T.~Cizmar and K.~Dholakia, ``Shaping the light transmission through a multimode
  optical fibre: complex transformation analysis and applications in
  biophotonics,'' {\em Optics Express}, vol.~19, no.~20, pp.~18871--18884,
  2011.

\bibitem{PapadopoulosOE2012}
I.~N. Papadopoulos, S.~Farahi, C.~Moser, and D.~Psaltis, ``Focusing and
  scanning light through a multimode optical fiber using digital phase
  conjugation,'' {\em Opt. Express}, vol.~20, no.~10, pp.~10583--10590, 2012.

\bibitem{ChoiPRL2012}
Y.~Choi, C.~Yoon, M.~Kim, T.~D. Yang, C.~Fang-Yen, R.~R. Dasari, K.~J. Lee, and
  W.~Choi, ``Scanner-free and wide-field endoscopic imaging by using a single
  multimode optical fiber,'' {\em Phys. Rev. Lett.}, vol.~109, no.~20,
  p.~203901, 2012.

\bibitem{AndresenOL2013}
E.~R. Andresen, G.~Bouwmans, S.~Monneret, and H.~Rigneault, ``Toward endoscopes
  with no distal optics: video-rate scanning microscopy through a fiber
  bundle,'' {\em Opt. Lett.}, vol.~38, no.~5, pp.~609--611, 2013.

\bibitem{AndresenOE2013}
E.~R. Andresen, G.~Bouwmans, S.~Monneret, and H.~Rigneault, ``Two-photon
  lensless endoscope,'' {\em Opt. Express}, vol.~21, no.~18, pp.~20713--20721,
  2013.

\bibitem{OhayonBOE2018}
S.~Ohayon, A.~M. Caravaca-Aguirre, R.~Piestun, and J.~J. DiCarlo, ``Minimally
  invasive multimode optical fiber microendoscope for deep brain fluorescence
  imaging,'' {\em Biomed. Opt. Express}, vol.~9, no.~4, pp.~1492--1509, 2018.

\bibitem{Vasquez-LopezLSA2018}
S.~Vasquez-Lopez, R.~Turcotte, V.~Koren, M.~Pl\"{o}schner, Z.~Padamsey, M.~J.
  Booth, T.~Cizmar, and N.~Emptage, ``Subcellular spatial resolution achieved
  for deep-brain imaging in vivo using a minimally invasive multimode fiber,''
  {\em Light: Science \& Applications}, vol.~7, p.~110, 2018.

\bibitem{LoterieOE2015}
D.~Loterie, S.~Farahi, I.~Papadopoulos, A.~Goy, D.~Psaltis, and C.~Moser,
  ``Digital confocal microscopy through a multimode fiber,'' {\em Opt.
  Express}, vol.~23, no.~18, pp.~23845--23858, 2015.

\bibitem{TsvirkunOL2016}
V.~Tsvirkun, S.~Sivankutty, E.~R. Andresen, and H.~Rigneault, ``Wide-field
  lensless endoscopy with multi-core fiber,'' {\em Opt. Lett.}, vol.~41,
  no.~20, pp.~4771--4774, 2016.

\bibitem{PopoffPRL2010}
S.~M. Popoff, G.~Lerosey, R.~Carminati, M.~Fink, A.~C. Boccara, and S.~Gigan,
  ``Measuring the transmission matrix in optics: An approach to the study and
  control of light propagation in disordered media,'' {\em Phys. Rev. Lett.},
  vol.~104, no.~10, p.~100601, 2010.

\bibitem{PloschnerNatPhoton2015}
M.~Pl\"{o}schner, T.~Tyc, and T.~Cizmar, ``Seeing through chaos in multimode
  fibres,'' {\em Nat. Photon.}, vol.~9, pp.~529--535, 2015.

\bibitem{FarahiOE2013}
S.~Farahi, D.~Ziegler, I.~N. Papadopoulos, D.~Psaltis, and C.~Moser, ``Dynamic
  bending compensation while focusing through a multimode fiber,'' {\em Opt.
  Express}, vol.~21, no.~19, pp.~22504--22514, 2013.

\bibitem{Caravaca-AguirreOE2013}
A.~M. Caravaca-Aguirre, E.~Niv, D.~B. Conkey, and R.~Piestun, ``Real-time
  resilient focusing through a bending multimode fiber,'' {\em Opt. Express},
  vol.~21, no.~10, pp.~12881--12888, 2013.

\bibitem{AndresenJBO2016}
E.~R. Andresen, S.~Sivankutty, V.~Tsvirkun, G.~Bouwmans, and H.~Rigneault,
  ``Ultra-thin endoscopes based on multi-core fibers and adaptive optics: a
  status review and perspectives,'' {\em J. Biomed. Opt.}, vol.~21, no.~12,
  p.~121506, 2016.

\bibitem{TsvirkunOE2017}
V.~Tsvirkun, S.~Sivankutty, G.~Bouwmans, O.~Vanvincq, E.~R. Andresen, and
  H.~Rigneault, ``Bending-induced inter-core group delays in multicore
  fibers,'' {\em Opt. Express}, vol.~25, no.~25, pp.~31863--31875, 2017.

\bibitem{AndresenJOSAB2015}
E.~R. Andresen, S.~Sivankutty, G.~Bouwmans, L.~Gallais, S.~Monneret, and
  H.~Rigneault, ``Measurement and compensation of residual group delay in a
  multi-core fiber for lensless endoscopy,'' {\em J. Opt. Soc. Am. B}, vol.~32,
  no.~6, pp.~1221--1228, 2015.

\bibitem{SivankuttyOL2016}
S.~Sivankutty, E.~R. Andresen, G.~Bouwmans, T.~G. Brown, M.~A. Alonso, and
  H.~Rigneault, ``Single shot polarimetry imaging of multicore fiber,'' {\em
  Opt. Lett.}, vol.~41, no.~9, pp.~2105--2108, 2016.

\bibitem{SchermerJQE2007}
R.~T. Schermer and J.~H. Cole, ``Improved bend loss formula verified for
  optical fiber by simulation and experiment,'' {\em IEEE J. Quantum
  Electron.}, vol.~43, no.~10, pp.~899--909, 2007.

\bibitem{SnyderLove}
A.~W. Snyder and J.~D. Love, {\em Optical Waveguide Theory}.
\newblock Chapman \& Hall.

\bibitem{Nicolet2007}
A.~Nicolet, F.~Zolla, Y.~O. Agha, and S.~Guenneau, ``Leaky modes in twisted
  microstructured optical fibers,'' {\em Wave Random Complex}, vol.~17, no.~4,
  pp.~559--570, 2007.

\bibitem{NapiorkowskiOE2014}
M.~Napiorkowski and W.~Urbanczyk, ``Rigorous simulations of a helical core
  fiber by the use of transformation optics formalism,'' {\em Opt. Express},
  vol.~22, no.~19, pp.~23108--23120, 2014.

\end{thebibliography}
\end{document}